\documentclass{aastex}
\usepackage{emulateapj5,epsfig}

\newenvironment{inlinefigure}{ 
\def\@captype{figure} 
\noindent\begin{minipage}{0.999\linewidth}\begin{center}} 
{\end{center}\end{minipage}\smallskip} 
\newcommand{\beq}{\begin{equation}}
\newcommand{\eeq}{\end{equation}}

\font\tenbg=cmmib10 at 10pt

\def \rvecphi{{\hbox{\tenbg\char'036}}}

\def \rvecmu{{\hbox{\tenbg\char'026}}}

\def \rvecOmega {{\hbox {\tenbg\char'012}}}

\begin{document}
\title{Torsional Magnetic Oscillations in
Type I X-Ray Bursts}

\author{
R.V.E. Lovelace\altaffilmark{1}, A.K. Kulkarni\altaffilmark{2},
and  M.M. Romanova\altaffilmark{2}}

\altaffiltext{1}{Departments of Astronomy and  Applied and
Engineering Physics,
Cornell University, Ithaca, NY 14853-6801;
RVL1@cornell.edu}

\altaffiltext{2}{Department of Astronomy,
Cornell University, Ithaca, NY 14853-6801; akk26@cornell.edu,
romanova@astro.cornell.edu}

\begin{abstract}

  Thermonuclear burning on the surface of a neutron
star causes the  expansion of a thin outer layer
of the star, $\Delta R(t)$.
  The layer rotates slower
than the star due to angular momentum conservation.
  The shear between the star and
the layer acts to twist the star's dipole magnetic field giving
at first a trailing spiral field.
   The twist of the field acts
in turn to `torque up' the layer
increasing its specific angular momentum.
  As the layer cools and contracts, its excess  specific
angular momentum causes it to {\it rotate faster} than the
star which gives a leading spiral magnetic field.
  The process repeats, giving rise to torsional
oscillations.
    We derive equations for the angular velocity
and magnetic field of the layer taking into
account the diffusivity and viscosity which are probably due
to turbulence.
   The magnetic field causes a nonuniformity of
the star's photosphere (at the top
of the heated layer),  and this gives
rise to the observed X-ray oscillations.
   The fact that the layer periodically rotates faster
than the star means that the X-ray oscillation frequency
may ``overshoot'' the star's rotation frequency.
Comparison of the theory is
made with observations of Chakrabarty et al. (2003)
of an X-ray burst of SAX J1804.4-3658.

\end{abstract}
\section{Introduction}

Type I X-ray bursts have been observed in a
number of low-mass X-ray binary systems.
They are characterized by a rapid
($\sim 1-10$ s) rise in the observed
flux, followed by a relatively slow
($\sim 10-100$ s) decline
(see reviews by Lewin et al. 1995;
Strohmayer \& Bildsten 2003).
  Some objects show highly coherent
oscillations during the burst.
The oscillation frequency varies slightly
during the burst, but asymptotically
approaches the neutron star rotation
frequency (Strohmayer \& Bildsten 2003
and references therein; van der Klis 2000).
The oscillations indicate
nonuniform emission from the  star's
photosphere.
surface of the star.
  The nonuniformity is likely  due to
the star's non-aligned dipole magnetic field.

It is generally thought that Type I
X-ray  bursts are caused by thermonuclear
explosions on the surface of a neutron star.
  If the neutron star's magnetic field
is strong enough to channel accretion
onto the star's surface, then the
thermonuclear burning will not
be uniform over the surface.
  It is then likely that the
oscillations are produced
by spin modulation of the burst flux.
  The exact mechanism of the oscillations,
however, remains unclear in spite of a
number of studies (Cumming
et al. 2002; Spitkovsky et al. 2002;
Bhattacharya et al. 2005; Piro \&
Bildsten 2005).
   A significant
difficulty encountered by models is
explaining the oscillation frequency drifts.
   Cumming et al. (2002) investigated
the effect of hydrostatic expansion of the
neutron star atmosphere on the burst
oscillations.
   An expanded layer will
rotate slower than the star in order to
conserve angular momentum, and then return
to the stellar rotation rate
during the contraction phase.
 This could cause the drift in the oscillation
frequency.
  However, Cumming et al.
concluded that this phenomenon alone is
not enough to explain the largest observed
frequency drifts.
  They also noted that a strong
stellar magnetic field could inhibit
angular velocity changes in the atmosphere.
  In some bursts the oscillation frequency
`overshoots' its asymptotic frequency
(Strohmayer 1999; Miller 2000; Chakrabarty et al. 2003).

Here, we investigate the influence of
a star's magnetic field on the rotation
of its  heated outer layer.
  The theory is developed in \S 2 where
the induction equation is combined with
angular momentum conservation to obtain
equations for the rotation rate and magnetic
field in the heated layer.
  Later in \S 2 we include the
influence of the magnetic diffusivity
and the viscosity of the heated layer.
   In \S 3 we discuss sample solutions for
the rotation rate and magnetic field in
the layer for a case relevant to a burst
of SAX J1808.4-3658 described by Chakrabarty et
al. (2003).
  Section 4 gives the conclusions of this
work.

\section{Theory}

     We consider that a thermonuclear explosion on
the surface of the star creates a heated layer of
mass $\Delta M={\rm const}$ and
thickness $\Delta R(t)$ outside of the
star's equilibrium radius $R_*$, with
  $\Delta R$ considered for simplicity to be
uniform over the star's surface.
   The actual layer may be an equatorial
band (Inogamov \& Sunyaev 1999).
  The heating of the layer occurs rapidly and
it expands rapidly giving
$\Delta R(0)$.
    Subsequently, the layer slowly cools and
$\Delta R(t)$ evolves and then decreases
on a time-scale $10-100$ s.
  We consider that $\Delta R(t)$ is a known function
of time obtained from  the radiative
transfer  and the radial force balance.

    We use a spherical inertial
coordinate system $(R,\theta,\phi)$
and a coordinate system $(R,\theta,\phi^\prime)$ rotating with
the star at the constant angular rate $\Omega_*$.
   The flow field in the layer is
\begin{equation}
{\bf v}=v_R\hat{\bf R} + v_\phi \hat{\rvecphi~}~.
\end{equation}
 For the radial velocity we assume
\begin{equation}
v_R={R-R_* \over \Delta R(t)}~ \stackrel{\bullet}{\Delta R}(t)~,
\end{equation}
where $\stackrel{\bullet}{\Delta R}\equiv d\Delta R/dt$.
  The azimuthal velocity can be written
as
\begin{equation}
v_\phi =R_*\Omega_*\sin\theta \hat{\rvecphi~}
+v_\phi^\prime \hat{\rvecphi~}~,
\end{equation}
where $v_\phi^\prime(R_*,\theta)=0$.

   We consider the case of an orthogonal rotator
where the magnetic moment $\rvecmu$ is perpendicular
to the rotation axis $\rvecOmega_*$ as sketched
in Figure 1.
  This situation would give X-ray oscillations
at about twice the rotation frequency of the star.
 This case is valuable in that  it is amenable
to an analytic treatment, and it indicates
the behavior for non-orthogonal rotators
where the X-ray oscillations are at about the rotation
frequency of the star.
   For a non-orthogonal rotator, we suggest
that the value of $\mu$ in the following
expressions be replaced by $\mu_\perp =
|\rvecmu \times \rvecOmega|/|\rvecOmega|$.

    We focus on the equatorial
region of the star's surface,
$|\theta -\pi/2| \leq \pi/6$.
    The evolution of the magnetic field
within the layer
is described by the induction equation
which follows from Faraday's law and
infinite conductivity,
\begin{equation}
{\partial {\bf B} \over \partial t}
={\bf \nabla \times}({\bf v \times B})~,
\end{equation}
with ${\bf v}$ given by equation (1).
Later, we include a finite conductivity.
Inside the star $R\leq R_*$ the magnetic
field is frozen-in and rotates with the star.
   Outside the layer, $R > R_*+\Delta R$,
the magnetic field is considered to be
a vacuum field which rotates but is
otherwise unaffected by processes in the layer.
  Any   field lines linking the star and disk
are expected to be opened owing the large
difference between the angular velocity of
the inner disk and the star (Lovelace,
Romanova, \& Bisnovatyi-Kogan 1995).

   It is useful to describe the magnetic
field in terms of a flux function
$\Psi \equiv R A_\theta$,  where ${\bf A}=(0,A_\theta,0)$
is the vector potential.
Therefore, in the equatorial region of the star,
\begin{equation}
B_R =-~{1\over R^2}{\partial \Psi \over \partial \phi}
~,\quad \quad B_\phi =
{1\over R}{\partial \Psi \over \partial R}~,
\end{equation}
and $B_\theta=0$.
    We  adopt a reference frame rotating with
the star with $\phi^\prime=\phi-\Omega_*t$
the azimuth relative to a fixed point on the star.
In place of equation (4) we have
\begin{equation}
{\partial \Psi \over \partial t}=
-({\bf v^\prime \cdot \nabla})\Psi =
- v_R{\partial \Psi \over \partial R}
-{v_\phi^\prime \over R}{\partial \Psi
\over \partial \phi^\prime}~,
\end{equation}
where ${\bf v}^\prime=(v_R,0, v_\phi^\prime)$
and $v_\phi^\prime =v_\phi-\Omega_*R_*$.
This equation  simply says that the
flux function $\Psi$ is advected with the flow.

\begin{inlinefigure}
\centerline{\epsfig{file=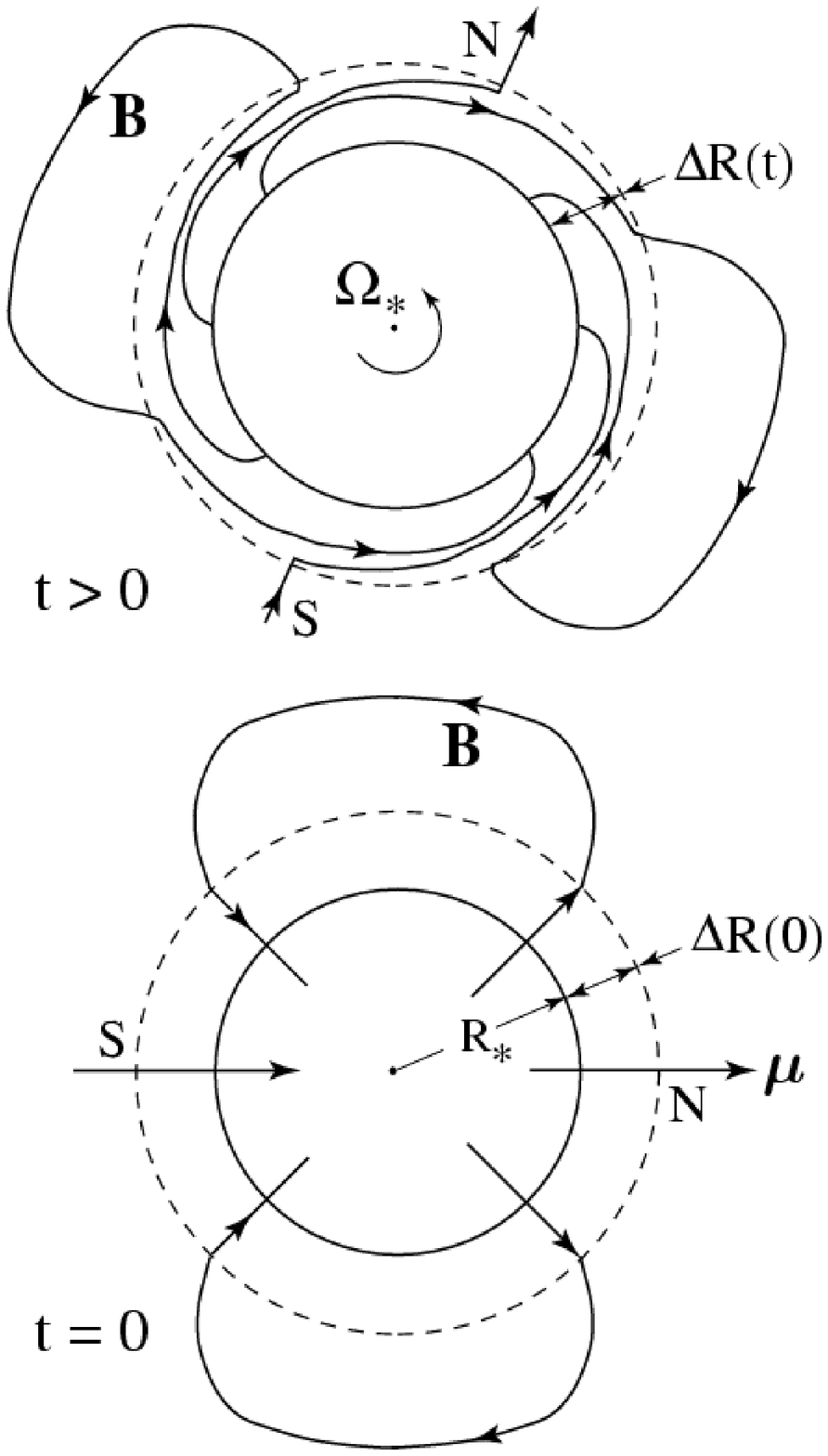,
height=4.in,width=2.4in}}
\epsscale{0.8}
\figcaption{The bottom panel shows the initial
configuration and the top panel shows the
configuration after some  time.  $\Delta R(t)$
is the thickness of the heated layer,
and $N$ and $S$ indicate the north and south
magnetic poles of the star.
The vacuum field outside of the layer
($R>R_*+\Delta R$) rotates
but does not change its form.}
\end{inlinefigure}

   We solve equation (6) by taking
\begin{equation}
\Psi(R,\phi^\prime,t)=iB_0R_*^2
\exp\bigg[i\int_{R_*}^R dR^\prime k_R(R^\prime,t)
+i\phi^\prime\bigg]~.
\end{equation}
The physical solution is the real part of
$\Psi$, denoted $\Re(\Psi)$;
$B_0$ is a constant field strength at
the magnetic poles of the star; and $k_R$
is the radial wavenumber which remains to
be determined.
  This form of solution was proposed earlier
for magnetized disks (Bisnovatyi-Kogan \& Lovelace 2000).
    From equation (5) we have
\begin{equation}
B_R =B_0{R_*^2 \over R^2}~\exp(...)~,\quad\quad
B_\phi=-B_0{k_R R_*^2 \over R}~\exp(...)~,
\end{equation}
where the exponential factors are the same
as in equation (7).
    Initially, there is no toroidal magnetic
field so that $k_R(t=0)=0$.

    Substitution of equation (7) into (6) gives
\begin{equation}
\int_{R_*}^R dR^\prime {\partial k_R \over \partial t}
=-k_R{ R-R_* \over \Delta R} \stackrel{\bullet}{\Delta R}
-{v_\phi^\prime \over R}~.
\end{equation}
Taking the $R-$derivative of this equation
gives
\begin{equation}
{\partial k_R \over \partial t}=
-~{k_R \stackrel{\bullet}{\Delta R} \over \Delta R}
-{R-R_* \over \Delta R} \stackrel{\bullet}{\Delta R} {\partial k_R \over
\partial R} -{\partial \over \partial R}\left({v_\phi^\prime \over
R}\right)~.
\end{equation}
In the following we assume that
$|(\Delta R/k_R)(\partial k_R/\partial R)| \ll 1$.
As a result the middle term of the right-hand side
of this equation can be neglected.

   Note that $\delta \omega(R,t) \equiv v_\phi^\prime/R$
is the difference between the angular velocity of the
layer and the star.   We let $\ell$ be the specific
angular momentum of the layer matter and $\ell_*=R_*^2\Omega_*$
the specific angular momentum of matter in
the equatorial region of the star.
   We then have
\begin{equation}
\delta \ell =\ell-\ell_* = 2(R-R_*)R_*\Omega_*
+ R_*^2 \delta \omega~.
\end{equation}
   We assume the factorable dependence
$\delta \ell =R_*\Omega_*(R-R_*){\cal L}(t)$,
where ${\cal L}(t)$ is a dimensionless function
determined subsequently.
 Consequently,
\begin{equation}
\delta \omega(R,t) =\Omega_*{R-R_* \over R_*}
\big[{\cal L}(t)-2\big]~.
\end{equation}
Therefore, equation (10) becomes
$$
{d k_R \over dt} = -{k_R \over \Delta R}~\stackrel{\bullet}{\Delta R}
-{\Omega_* \over R_*}\big[{\cal L}(t)-2\big]~,
$$
or more compactly,
\begin{equation}
{d (k_R \Delta R) \over dt}=-\Omega_* {\Delta R \over R_*}~
\big[{\cal L}(t)-2\big]~.
\end{equation}
Introducing
\begin{equation}
\Delta \omega(t) \equiv \Omega_*{\Delta R(t)\over R_*}
\big[{\cal L}(t)-2\big]~,
\end{equation}
 allows us to rewrite this equation as
\begin{equation}
{d (k_R \Delta R) \over dt}=-\Delta \omega~.
\end{equation}
Because $k_R \propto -B_\phi(R_*)$ from equation (8),
$k_R \Delta R$ is proportional to the toroidal
flux (in one direction) in the layer.
   Equation (15) expresses the fact that the rate
of change of this flux is proportional to the
`rate of twisting,'  $\Delta \omega(t)$.

\subsection{Angular Momentum Conservation}

The total angular momentum of the heated layer minus
$\Delta M \Omega_*R_*^2$ is simply
\begin{eqnarray}
\Delta L&=&R_*\Omega_*
\int_{R_*}^{R_*+\Delta R} dR (R-R_*){\cal L}(t) {dM
\over dR}~,
\nonumber\\
&=&c_MR_*\Omega_* \Delta M \Delta R(t){\cal L}(t)~.
\end{eqnarray}
Here, $c_M$ is a dimensionless constant of order
unity assuming a self-similar mass distribution
in the layer, $dM/dR =f[(R-R_*)/\Delta R]$.

Changes in $\Delta L$ are due to twisting of the
magnetic field.
  The vacuum field in the region $R>R_*+\Delta R$
rotates but it is otherwise
unaffected by processes within the heated layer.
   Thus the change in $\Delta L$ is due to the
outflow of angular momentum from the
surface of the star $R=R_*$.  That is,
\begin{eqnarray}
{d \Delta L \over dt}&=&- \int dS_R~R\sin\theta ~
{\Re(B_R) \Re(B_\phi) \over 4\pi}\bigg|_{R=R_*}~,
\nonumber\\
&=&~{1\over 4}k_R R_*^4 B_0^2~,
\end{eqnarray}
where  the fields are given by equation (8)
and the integration is over the area $2\pi R_*^2$
of the equatorial region of the star.
Using equation (16) gives
\begin{equation}
{d \over dt}\left({\cal R}{\cal L}\right)=
{\omega_B^2 \over \Omega_*}~{\cal K}~,
\end{equation}
where
$$
{\cal R}\equiv {\Delta R \over R_*}~,
\quad {\cal K} \equiv k_R R_*~,\quad
\omega_B^2 \equiv {R_*B_0^2 \over 4c_M \Delta M}~,
$$
where $\omega_B={\rm const}$ is an
Alfv\'en frequency  of the heated layer.
The actual oscillation frequency of a layer
of constant thickness $\Delta R$ is
larger than $\omega_B$ by a factor of $(R_*/\Delta R)^{1/2}$
(see equation 21).

  For representative values from Cumming and Bildsten (2000),
\begin{equation}
\omega_B \approx
{3.16/{\rm s} \over \sqrt{4 c_M}}
\left({B_0 \over 10^8{\rm G}}\right)
\times \left({10^{21}{\rm g} \over \Delta M}\right)^{1/2}
\left({R_* \over 10^6{\rm cm}}\right)^{1/2}~.
\end{equation}
    For a layer of constant thickness, the Alfv\'en
speed in the layer is
$v_A \approx (\omega_B/\pi)\sqrt{R_*\Delta R}$.
For the given reference values and $\Delta R=10^3{\rm cm}$,
$v_A \approx 3.2\times 10^4$cm/s.

\subsection{Main Equations without Dissipation}

Using equation (14), equations (15) and (18) can be written as
\begin{eqnarray}
{d({\cal R}{\cal K}) \over dt}&=&- \Omega_* {\cal W}~,
\nonumber \\
{d {\cal W} \over dt}&=&{\omega_B^2 \over \Omega_*}~{\cal K}
-2\stackrel{\bullet}{\cal R}~,
\end{eqnarray}
where
$$
{\cal W}\equiv {\Delta \omega \over \Omega_*}~,
$$
is the dimensionless `rate of twisting.'
  The initial conditions are that ${\cal K}(0)=0$ which
corresponds to no initial toroidal magnetic field,
and ${\cal W}(0)=-2{\cal R}(0)$ which corresponds to
the specific angular momentum of the layer  being
equal to its equilibrium value (i.e., ${\cal L}(0)=0$).

   Equations (20) are a linear system for $({\cal K},~{\cal W})$
in that ${\cal R}(t)$ is a given function.
  For ${\cal R}={\rm const}$, the solution of equations (20)
is oscillatory with angular frequency
\begin{equation}
\omega_{osc}={\omega_B \over \sqrt{\cal R}}~.
\end{equation}
Thus the period of the oscillation is proportional
to $\sqrt{\Delta R}$.

  Equations (18) have the form
of Hamilton's equations,
$\stackrel{\bullet}{Q}=\partial{\cal H}/\partial P$,
$\stackrel{\bullet}{P}= -\partial{\cal H}/\partial Q$, with
$P \equiv \omega_B^2{\cal K}{\cal R}/\Omega_*$ the
canonical momentum,
$Q \equiv {\cal W}$ the canonical coordinate, and the Hamiltonian
${\cal H}=  P^2/(2{\cal R})+\omega_B^2Q^2/2
-2\stackrel{\bullet}{\cal R}P$.
   These equations are idealized
in the respect that they neglect the magnetic
diffusivity and the viscosity of the layer.

\begin{inlinefigure}
\centerline{\epsfig{file=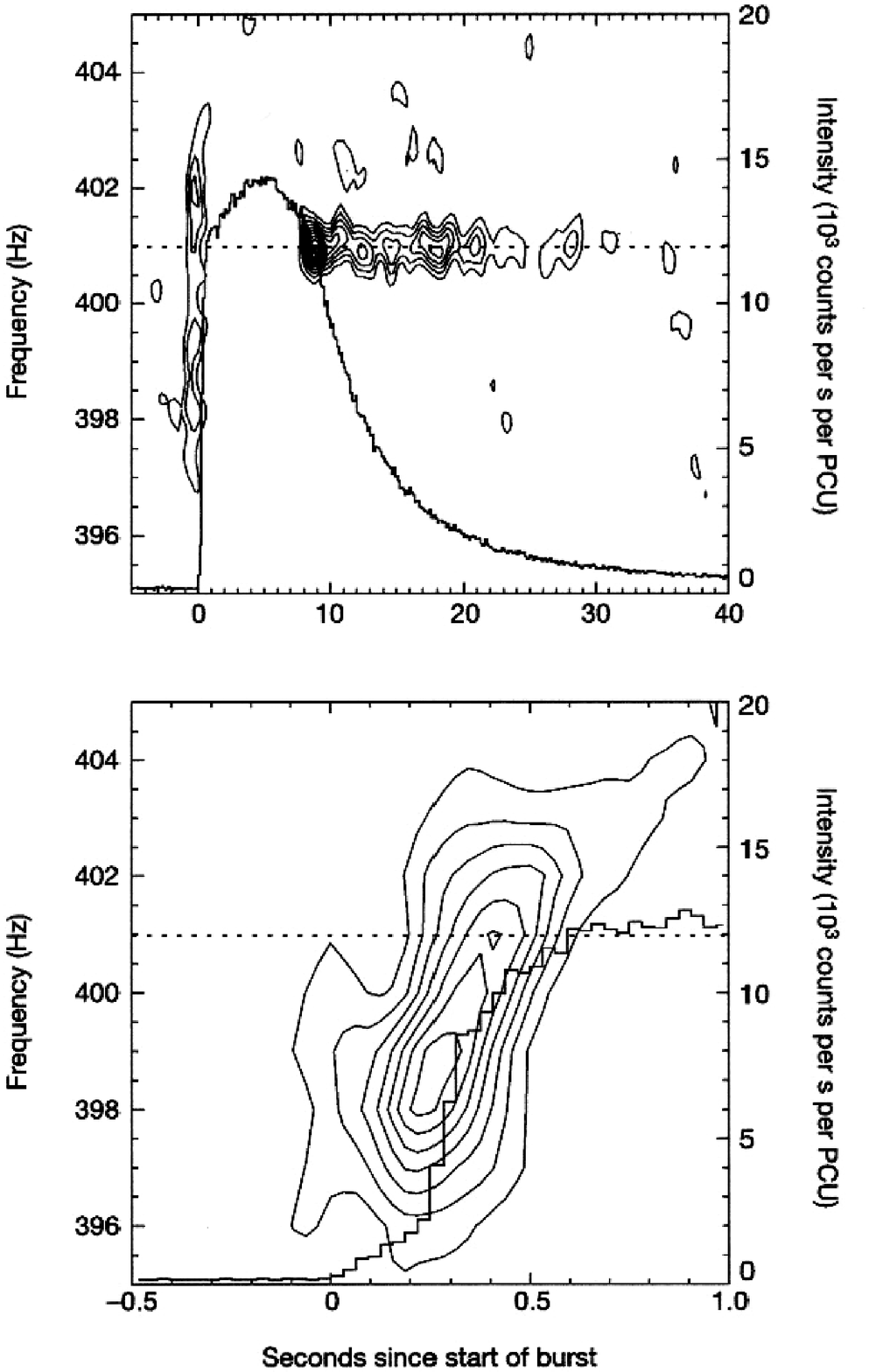,
height=5.in,width=3.2in}}
\epsscale{0.8}
\figcaption{The contours show dynamic power spectra
of the X-ray burst of SAX J1808.4-3658 on October
18, 2000 (Figure 1 of Chakrabarty et al. 2003).
The solid curves show the  X-ray count rate.
The bottom panel shows a close in view
of the top panel.
}
\end{inlinefigure}

\subsection{Equations with Magnetic Diffusivity}

    For finite conductivity $\sigma$
of the layer plasma,  the term $\eta_m \nabla^2{\bf B}$
is added to the right-hand side of equation (4),
where $\eta_m=c^2/(4\pi \sigma)$ is the magnetic
diffusivity.   In place of equation (6) we
have
\begin{equation}
{\partial \Psi \over \partial t}=
-({\bf v^\prime \cdot \nabla})\Psi +\eta_m \Delta^* \Psi~,
\end{equation}
where $\Delta^* \equiv \partial^2/\partial R^2
+(1/R^2)\partial^2/\partial \phi^2$ is the adjoint
Laplacian operator.

    We solve equation (22) with a generalization
of equation (7) where the radial wavenumber
is complex, $k_R\rightarrow k_{Rr}+ik_{Ri}$.
We introduce the dimensionless
wavenumber $({\cal K}_r, {\cal K}_i) \equiv
(k_{Rr}R_*,~k_{Ri}R_*)$, and $\tau_m \equiv R_*^2/\eta_m$.
    Thus
\begin{eqnarray}
{d({\cal R}{\cal K}_r) \over dt}&=&-~ \Omega_* {\cal W}
-{2\over \tau_m}{\cal K}_r{\cal K}_i ~,
\nonumber \\
{d({\cal R}{\cal K}_i) \over dt}&=&
-~{1\over \tau_m}({\cal K}_r^2+1-{\cal K}_i^2)~,
\nonumber\\
{d {\cal W} \over dt}&=&{\omega_B^2 \over \Omega_*}~{\cal K}_r
-2\stackrel{\bullet}{\cal R}~.
\end{eqnarray}
The initial conditions are that ${\cal K}_r(0)=0$,
${\cal K}_i(0)=0$, and ${\cal W}(0)=-2{\cal R}(0)$.
   In contrast with equations (20), the evolution
equations with dissipation are nonlinear.

\subsection{Magnetic Diffusivity}

    Note that $\tau_m \equiv R_*^2/\eta_m$
is the Ohmic dissipation time-scale of the layer
multiplied by $(R_*/\Delta R)^2$.
   It is less than
or equal to its value given by the
classical Spitzer value of $\eta_m$,
\begin{equation}
\tau_m \leq \tau_{\rm Spitzer}\approx {10^{13}{\rm s} \over Z}
\left({R_* \over 10^6{\rm cm}}\right)^2
\left({T \over 10^9{\rm K}}\right)^{3/2}~,
\end{equation}
where $T$ is the temperature of the heated layer,
$Z$ is an average atomic charge, and the Coulomb
logarithm has been taken to be $8$
(Cumming \& Bildsten 2000).

    Instabilities and associated turbulence in the heated
layer are expected to give $\tau_m \ll \tau_{\rm Spitzer}$.
  Under different conditions the deep part of the  layer
may be convective and the top part radiative as discussed by
Cumming and Bildsten (2000).
  The sound speed in the heated layer $c_s \approx 2.9\times
10^8{\rm cm/s}(T/10^9{\rm K})^{1/2}$
is generally much larger than the Alfv\'en speed $v_A$
estimated previously as $v_A \sim 3.2\times 10^4$cm/s
at the beginning of the burst.
  Later in the burst the magnetic field is
twisted and the magnitude of the toroidal
field  increases substantially
due mainly to  the field being confined to the
thin layer $\Delta R$.
  Thus the Alfv\'en
crossing time $t_A = \Delta R/v_A$ decreases
from its initial value $\sim 0.03$s assuming
$\Delta R=10^3$ cm.
   The strong toroidal field may lead to a
buoyancy instability.
   Notice however that the layer has a strong
shear with a velocity difference across it
of  order  $\Delta v \approx 2\Omega_*\Delta R
\sim 5\times 10^6$cm/s so that the
shearing time  $t_{\rm shear} =\Delta R/\Delta v
\sim 2\times 10^{-4}$s is shorter than $t_A$
in the initial  part of  a burst.
   Furthermore, notice that the upward
buoyant motion of a blob
will be strongly influenced by
the Coriolis force $2\rho{\bf v \times }\rvecOmega$,
where ${\bf v}$ is the velocity in the rotating
reference frame.
   This force has a stabilizing affect on the
blob motion and tends to give circular `gyro'
motion of radius $r_g= |{\bf v}|/2\Omega\approx
6{~\rm cm~}[|{\bf v}|/(3\times 10^4{\rm cm/s})]$
 about the $\rvecOmega$ direction (Tritton 1988)
for $\Omega_*/2\pi =400$ Hz.
   Thus, $r_g$ for Alfv\'en
speed motions is smaller than
the layer thickness $\Delta R$ even for
conditions where the field strongly increased.

  Because $v_A \ll c_s$, the layer may also be unstable to
the magnetorotational instability
(Balbus \& Hawley 1998; Velikhov 1959;
Chandrasekhar 1960;  Moiseenko,
Bisnovatyi-Kogan, \& Ardeljan 2006).
   The saturation of the instabilities is
assumed to give rise to a
turbulent diffusivity $\eta_{mt}$ much larger than
the Spitzer value and a turbulent
viscosity $\nu_t \approx \eta_{mt}$
(Bisnovatyi-Kogan \& Ruzmaikin 1974).
The viscosity is taken into account in the next subsection.
  The turbulent viscosity
and diffusivity can be written in a form analogous
to the Shakura-Sunyaev (1973) ``alpha''
prescription for a thin disk
taking into account that
 the maximum eddy size is $\Delta R$ and the maximum
buoyant velocity is $\sim v_A$ with
$v_A$  the initial Alfv\'en speed.
  Later in a burst, $v_A$ increases substantially but
at the same time the mentioned effects of  shear
and Coriolis force  act to limit the motion
of buoyant blobs.
  Therefore, we estimate $\eta_{mt}=\alpha v_A \Delta R$
with $\alpha \leq 1$ which gives
\begin{equation}
\eta_{mt}\approx \nu_t
\approx 3\times 10^7 {{\rm cm}^2 \over {\rm s}}
~\alpha~
\left({\Delta R \over 10^3{\rm cm}}\right)
\left({v_A \over 3\times 10^4{\rm cm/s}}\right)~.
\end{equation}
 Note that for $\eta_m =3\times 10^7$cm$^2$/s, the
magnetic diffusion time across the layer is
$t_{\rm diff} = \Delta R^2/\eta_m \sim 0.033$s is
of the order of the initial Alfv\'en crossing time.

   We emphasize that
the values of $\eta_{mt}$ and $\nu_t$
are highly uncertain in that
relevant linear and nonlinear theory and simulations
have yet to be done.
  An extensive literature  exists on the
the theory and simulations of the related
problem of the stability and motion of magnetic flux
tubes in the convection zone and overshoot region
of the Sun (e.g., Sch\"usssler et al. 1994).
   It appears possible that these methods can
be adapted  to the magnetic
field stability of the X-ray burst sources.

\subsection{Equations with Diffusivity and Viscosity}

    Viscosity of the heated layer is taken into account
by including on the right-hand side of
equation (17) the viscous angular momentum flux across
the surface $R=R_*$.
  This term is
\begin{eqnarray}
\int dS_R R\sin\theta ~T^\nu_{R\phi} &=&
- \left[2\pi R_*^4\rho\nu {d\omega \over dR}\right]_{R=R_*}
\nonumber\\
&=&-2\pi R_*^4 \rho(R_*)~ \nu {\Delta \omega \over \Delta R}~,
\end{eqnarray}
where $T_{R\phi}^\nu $ is the viscous contribution to
the momentum flux-density tensor.
Thus, the equations for the layer including Ohmic and
viscous dissipation are
\begin{eqnarray}
{d({\cal R}{\cal K}_r) \over dt}&=&-~ \Omega_* {\cal W}
-{2\over \tau_m}{\cal K}_r{\cal K}_i ~,
\nonumber \\
{d({\cal R}{\cal K}_i) \over dt}&=&
-~{1\over \tau_m}({\cal K}_r^2+1-{\cal K}_i^2)~,
\nonumber\\
{d {\cal W} \over dt}&=&{\omega_B^2 \over \Omega_*}~{\cal K}_r
-{1 \over \tau_\nu}{{\cal W }\over {\cal R}^2}
-2\stackrel{\bullet}{\cal R}~ ,
\end{eqnarray}
where
\begin{equation}\tau_\nu=
c_\nu {R_*^2\over \nu}~.
\end{equation}
where $c_\nu \equiv
c_M \Delta M /[ 2\pi R_*^2\Delta R \rho(R_*)]$
is a dimensionless constant of order unity if
the mass distribution of the disk is self-similar.
  Numerical solutions of equations (26)
indicate that the influence of viscosity is
negligible for $\tau_\nu \gtrsim \tau_m$.

\section{Sample Solutions}

  To solve equations (27) we need
${\cal R}(t)=\Delta R(t)/R_*$ which is
not known from observations.
  A rough estimation of this function
can be made assuming  $\Delta R \propto T$
and the X-ray  intensity $I \propto T^4$.
  The X-ray intensity of the SAX J1808.4 burst shown
in Figure 2 (Chakrabarty et al. 2003) is fitted
approximately by $I\propto 1/\{1+[(t-5)/6.5]^2\}$
and for $10\le t \le 40$ accurately by $I\propto 1/t^2$.
The corresponding layer thickness is
${\cal R}={\cal R}(0)/\{1+[(t-5)/6.5]^2\}^{1/4}$,
with $t$ in seconds.

    Figure 3 shows a sample case relevant to
the observed burst shown in Figure 2.
   Figure 4 shows the trailing spiral
magnetic field  at $t=0.4$s for this case.
 Later, when ${\cal K}_r<0$, a leading spiral
field forms.
    The ratio $\beta \equiv B_\phi/B_R$ at $R=R_*$ is
$\beta=-{\cal K}_r$ and this is seen to vary from $-1450$
at $0.42$s to $420$ at $1.45$s.
    The  first positive peak of
$\Delta \omega/\Omega_*$ occurs at
$t_1 \approx 0.25/\omega_B$ for small
damping, $\tau_m \ge 10^5$s.
    The time between the first and second
positive peaks of $\Delta \omega/\Omega_*$
is $\Delta t_{12} \approx 0.53/\omega_B$.

   Note that $\tau_m =10^5$s corresponds
to a diffusivity $\eta_{mt}= R_*^2/\tau_m
= 10^7{\rm cm}^2/{\rm s}$ for $R_*=10^6$cm.
   From Figure 3 it is
seen that the twisting amplifies the
toroidal magnetic  field to a peak value of
the order ${\cal K}_r \approx 1400$
larger than the initial poloidal
field of the star.
   The amplification is due to the field being
confined to the thin layer $\Delta R$ rather
than it being wrapped many times around the  star.
  For the case of Figure 3, the field is wrapped
two turns in the clockwise
direction during the first $\approx 0.7$s,
and then it is unwrapped, wrapped, etc., subsequently.
  The magnetic diffusion time across the layer
is $t_{\rm diff}=\Delta R^2/\eta_m \approx 5.6$s.

   Figure 4 shows the ``phase slip,''
$\int_0^t dt^\prime \omega(t^\prime) - \Omega_* t$,
during a burst for the same conditions as Figure 3.
  Figure 5 shows the nature of the wrapped field
at $t=0.4$ s for the case of Figure 3.

   Figure 6 shows $\Delta \omega/\Omega_*$
over a longer time interval for
sample cases.
The period of the oscillations of $\Delta\omega$ is
proportional to $\sqrt{\Delta R}$ which
in this case is proportional to $1/t^{1/4}$.

\begin{inlinefigure}
\centerline{\epsfig{file=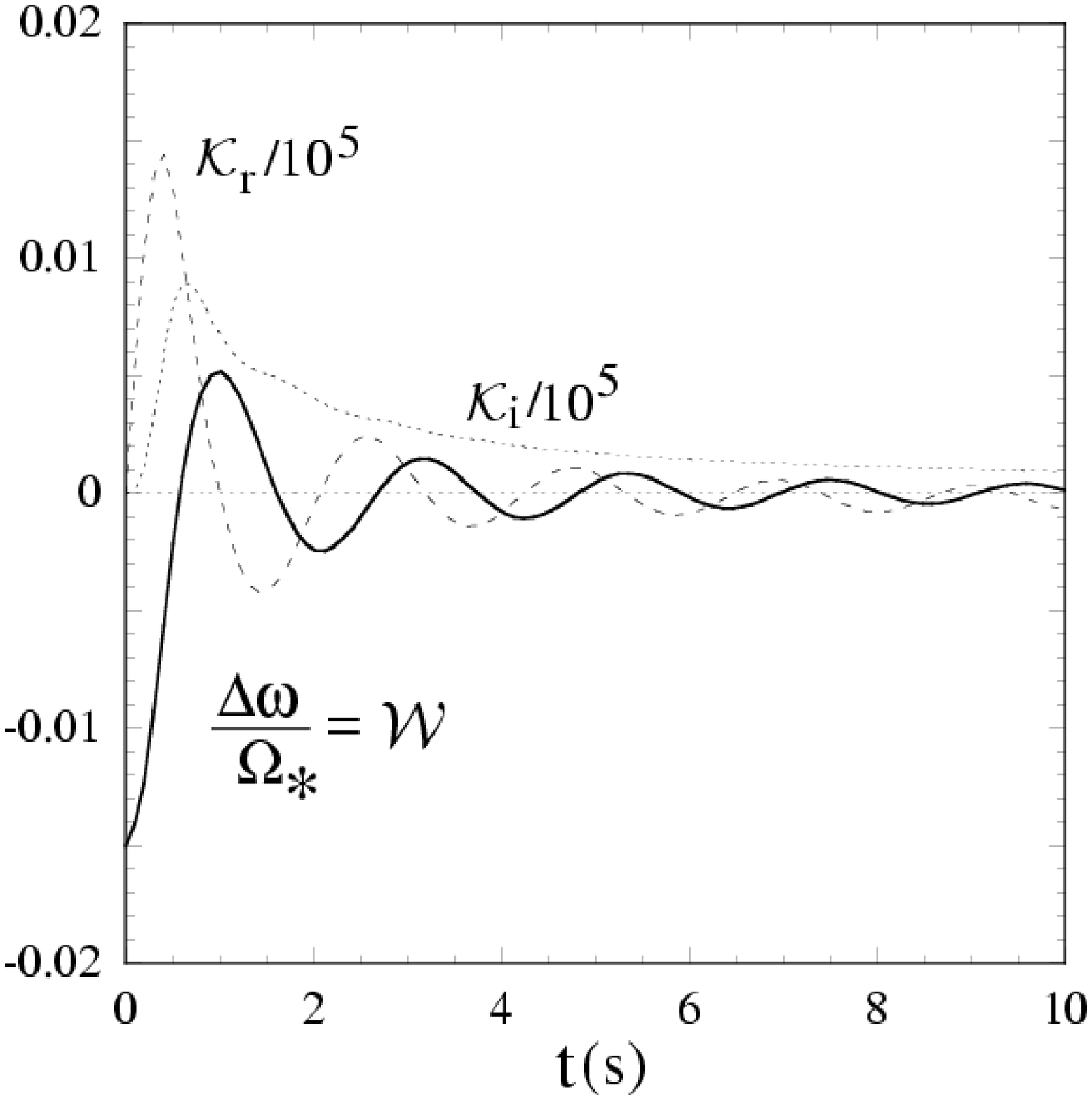,
height=3.2in,width=3.2in}}
\epsscale{0.8}
\figcaption{Sample solution of equations (23)
for ${\cal W}(t)=\Delta \omega/\Omega_*$, ${\cal K}_r(t)$
and ${\cal K}_i(t)$ relevant to the burst shown in
Figure 2.
   The conditions are
$\omega_B=0.252/{\rm s}$ and $\tau_m =10^5{\rm s}$,
which corresponds to a magnetic diffusion time across the layer
of $t_{\rm diff}=\Delta R^2/\eta_{m t} \approx 5.6$s.
Also, ${\cal R}(t)$ is given in the text with
${\cal R}(0)=0.0075$ and $\Delta \omega(0)/\Omega_*=0.015$.}
\end{inlinefigure}
\begin{inlinefigure}
\centerline{\epsfig{file=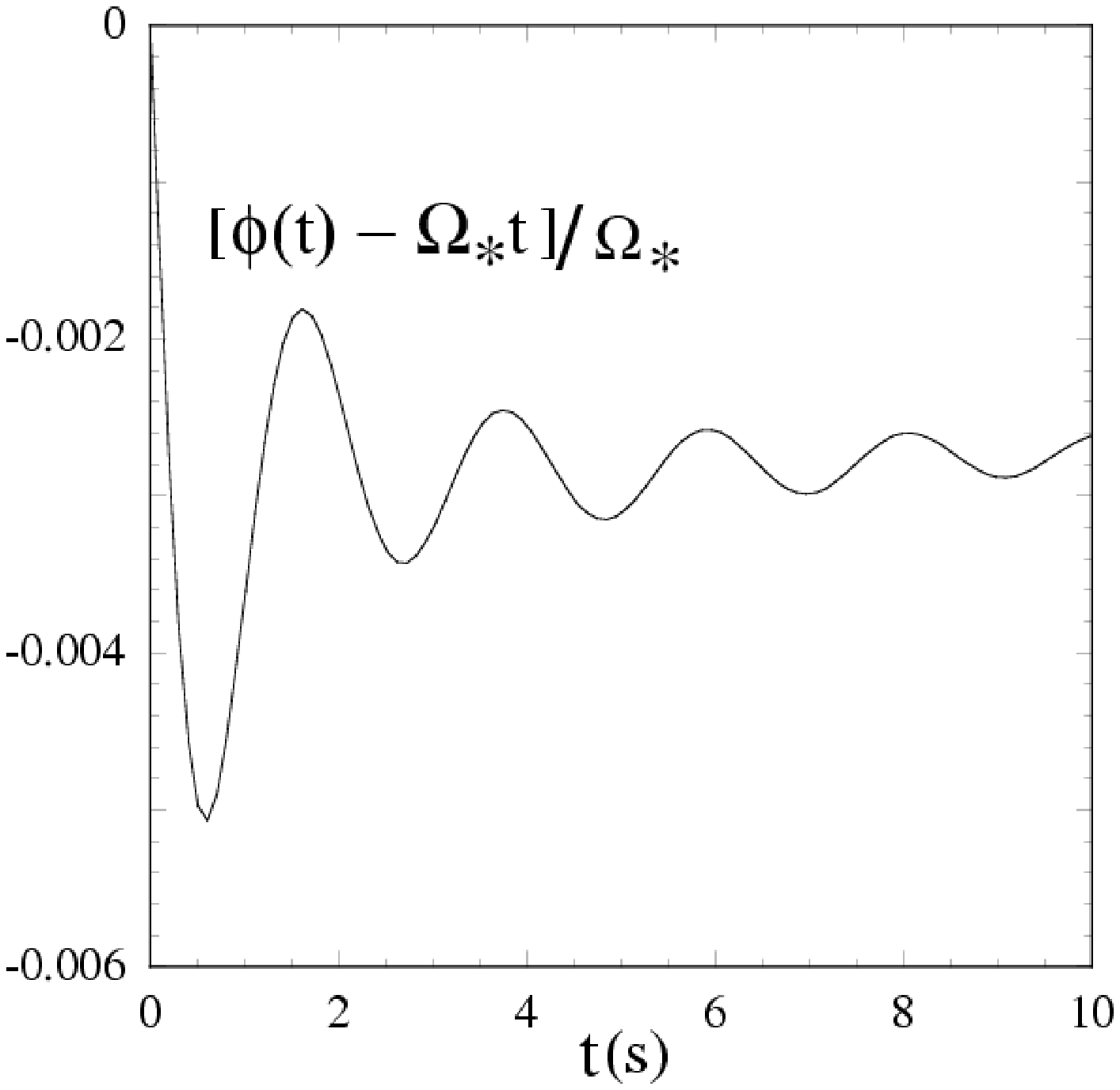,
height=3.2in,width=3.2in}}
\epsscale{0.8}
\figcaption{The ``phase-slip during a
burst for the
conditions of Figure 3, where
$\phi(t) =\int_0^t dt^\prime \omega(t^\prime)$.
}
\end{inlinefigure}

\begin{inlinefigure}
\centerline{\epsfig{file=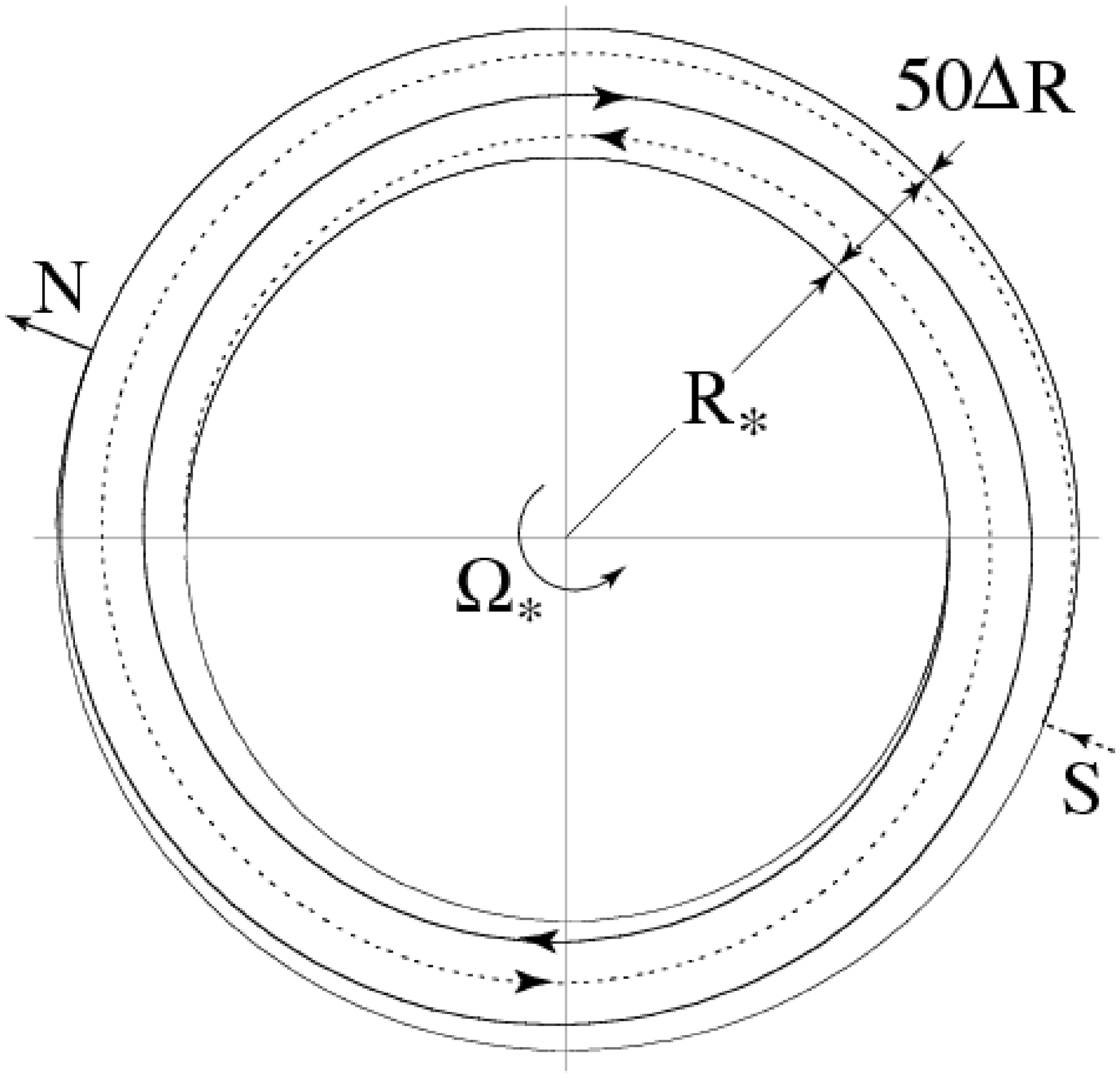,
height=2.4in,width=2.4in}}
\epsscale{0.8}
\figcaption{Field lines in the heated
layer at $t=0.4s$ for same
case as Figure 3.
  The radial thickness of
the layer has been expanded by a factor
of $50.$
   The field line is
given by $R_*\phi^\prime =-{\cal K}_r (R-R_*)
-({\cal K}_i/{\cal K}_r)R_*
\ln\{{\rm abs}[\cos({\cal K}_r(R-R_*)/R_*)]\}$,
where ${\cal K}_r=1447$, ${\cal K}_i=597.$, and
$\Delta R/R_*=0.00678$.
$N$ and $S$ indicate the north and south magnetic
poles.
 }
\end{inlinefigure}

\begin{inlinefigure}
\centerline{\epsfig{file=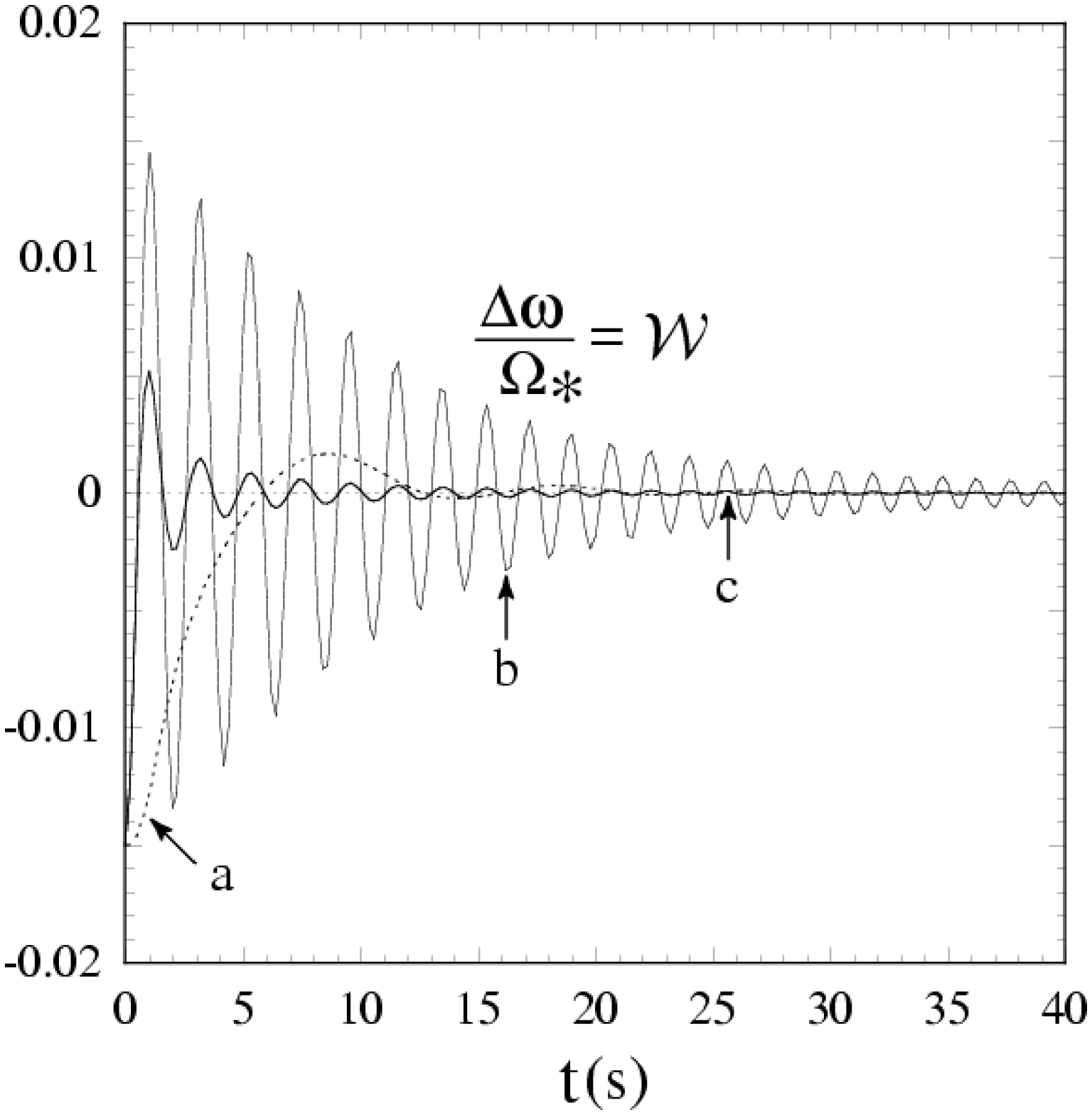,
height=3.2in,width=3.2in}}
\epsscale{0.8}
\figcaption{Sample solution of equations (23)
for ${\cal W}(t)=\Delta \omega/\Omega_*$
for a longer interval for ({\bf a}) $\omega_B=0.0504/{\rm s}$
and $\tau_m=10^6$s,
({\bf b}) $\omega_B=0.252/{\rm s}$,  $\tau_m=10^6$s,
and ({\bf c}) $\omega_B=0.252/{\rm s}$,  $\tau_m=10^5$s,
with other conditions the same as for Figure 3.
   Note that $\tau_m =10^5{\rm s}$
corresponds to a magnetic diffusion time across the layer
of $t_{\rm diff}=\Delta R^2/\eta_{m t} \approx 5.6$s.
  For $\tau_m =10^6{\rm s}$ the diffusion time is ten
times longer.
The period of the oscillations of $\Delta \omega$ decreases
gradually with time.}
\end{inlinefigure}

\section{Conclusions}

    We have derived a simple model for the
influence of a neutron star's magnetic field on the rotation
of its surface layer rapidly heated by thermonuclear burning.
   The model assumes that the star's magnetic moment
is perpendicular to its rotation axis.
   The burning
causes the  expansion of a thin outer layer
of the star, $\Delta R(t)$.
  The layer rotates slower
than the star due to angular momentum conservation.
  The shear between the star and
the layer acts to twist the star's magnetic field giving
at first a trailing spiral field.
  The twist of the field acts
in turn to `torque up' the layer
increasing its specific angular momentum.
  As the layer cools and contracts, its excess  specific
angular momentum causes it to {\it rotate faster} than the
star which gives a leading spiral magnetic field.
   The oscillation  of the angular velocity of the
layer is a result of the tension of the
twisted magnetic field.
  Non-uniformity of the star's photosphere (at the top
of the heated layer) is due the magnetic field and this gives
rise to the observed X-ray oscillations.
   The fact that the layer periodically rotates faster
than the star means that the X-ray oscillation frequency
may ``overshoot'' the star's rotation frequency.
   Observations by
Chakrabarty et al. (2003) of an X-ray burst
of SAX J1804.4-3658 show clear evidence of the
overshoot of the  frequency of the X-ray
oscillations.

    Equations of the model are for the
difference in angular velocity between the layer
and the star and the radial and toroidal
components of the magnetic field.
  The frequency of the oscillations is proportional
to the initial poloidal magnetic field of the
star, inversely proportional to the
square root of the mass of
the heated layer, and inversely proportional to
the square root of the layer's thickness.
   In the absence of magnetic diffusivity
and viscosity, the equations are linear
and constitute a Hamiltonian system.
  The equations become nonlinear with
the magnetic diffusivity included, but
the inclusion of viscosity adds a linear term.
  The diffusivity and viscosity are probably due
to turbulence in the heated layer but the
level of the turbulence is highly uncertain.
  The model has two important parameters, one is the
oscillation frequency proportional
to the initial magnetic field, and the other is the
damping time due to magnetic diffusivity.
   The value of the magnetic diffusivity is
highly uncertain.
   We find that the twisting can amplify the
toroidal magnetic  field to a peak value of
the order $10^3$ larger than the initial poloidal
field of the star.
   The amplification is due to the field being
confined to the thin layer $\Delta R$ rather
than it being wrapped many times around the  star.

\section*{Acknowledgements}
 We thank Dong Lai for stimulating discussions and
a referee for valuable comments.
This work was supported in part by NASA grants NAG5-13220,
NAG5-13060, by NSF grant AST-0507760.

\end{document}